\documentclass[a4paper,
               biblatex,     
               ]{jacow}
\usepackage{pdfpages,multirow,ragged2e} %

\makeatletter%
	\ifboolexpr{bool{xetex}}
	 {\renewcommand{\Gin@extensions}{.pdf,%
	                    .png,.jpg,.bmp,.pict,.tif,.psd,.mac,.sga,.tga,.gif,%
	                    .eps,.ps,%
	                    }}{}
\makeatother

\ifboolexpr{bool{xetex} or bool{luatex}} 
 {}                                      
 {\usepackage[utf8]{inputenc}}           

\usepackage[USenglish]{babel}
\usepackage{comment}
\ifboolexpr{bool{jacowbiblatex}}%
 {%
  \addbibresource{jacow-test.bib}
  \addbibresource{biblatex-examples.bib}
 }{}
\begin{document}

\title{Comparison of W\lowercase{arp}X and GUINEA-PIG for electron positron collisions \thanks{Work supported by the Director, Office of Science, Office of High Energy Physics of the U.S. Department of Energy (DOE) under Contract No. DE-AC02-05CH11231 and by the DOE under Contract DE- AC02-76SF005. 

}}

\author{B. Nguyen \thanks{william.nguyen21@imperial.ac.uk}, Imperial College London, London, United Kingdom \\
		A. Formenti, R. Lehe, J-L. Vay, Lawrence Berkeley National Laboratory, Berkeley, United States \\ 
  S. Gessner, SLAC National Accelerator Laboratory, Menlo Park, United States \\
  L. Fedeli, Universit\'e Paris-Saclay, CEA, LIDYL, 91191 Gif-sur-Yvette, France } 
\maketitle

\begin{abstract}
As part of the Snowmass'21 planning exercise, the Advanced Accelerator Concepts community proposed developing multi-TeV linear colliders and considered beam-beam effects for these machines. Such colliders operate under a high disruption regime with an enormous number of electron-positron pairs produced from QED effects. Thus, it requires a self-consistent treatment of the fields produced by the pairs, which is not implemented in state-of-the-art beam-beam codes such as GUINEA-PIG. WarpX is a parallel, open-source, and portable particle-in-cell code with an active developer community that models QED processes with photon and pair generation in relativistic laser-beam interactions. However, its application to beam-beam collisions has yet to be fully explored. In this work, we benchmark the luminosity spectra, photon spectra, and coherent production process from WarpX against GUINEA-PIG in the ILC and ultra-tight collision scenarios. Our performance comparison demonstrates a significant speed-up advantage of WarpX, ensuring a more robust and efficient modeling of electron-positron collisions at multi-TeV energies.
\end{abstract}

\section{Introduction}

The performance of a particle collider depends on two crucial parameters: the center-of-mass energy $\sqrt{s}$ and the luminosity $\mathcal{L}$ \cite{chao2008reviews, herr2006concept}. While the center-of-mass energy determines the upper limit of resolution strength, luminosity quantifies the number of events occurring per cross-section unit. As part of the Snowmass'21 planning exercise, the Advanced Accelerator Concepts (AAC) community proposed the development of multi-TeV linear colliders with luminosity up to $50 \times 10^{34} \ \mathrm{cm^{-2} s^{-1}}$ as the next-generation discovery machines \cite{barklow2023beam}. A key challenge of these machines is minimizing the detrimental beamstrahlung and background pair production effect while maintaining the target performance. Tackling this problem requires robust, accurate, and efficient simulation tools dedicated to beam-beam interaction studies. 

GUINEA-PIG, a state-of-the-art serial code developed at CERN, has been used for many beam-beam studies, informing the design of the International Linear Collider (ILC) and  Compact Linear Collider (CLIC) 
 \cite{schulte1999beam, rimbault2006incoherent, rimbault2007impact, shinton2011beam}. Despite its extensive use, GUINEA-PIG exhibits several limitations, such as the absence of a self-consistent treatment for pair generation, lack of parallelization, and outdated documentation and developer support. In contrast, WarpX is an exascale open-source particle-in-cell (PIC) code with strong performance, flexibility, up-to-date documentation and maintenance within an active, and multi-disciplinary community \cite{fedeli2022pushing}. In this paper, we benchmark the recently implemented beam-beam effect in WarpX against GUINEA-PIG with two test cases: established ILC designs and ultratight spherical beam collisions as outlined in references \cite{bambade2019international} and \cite{yakimenko2019prospect}.

\section{Theory}

Equation \eqref{eq:lumi} shows the geometric luminosity $\mathcal{L}$ from a head-on collision of two Gaussian beams. The transverse sizes ($\sigma_x,  \sigma_y)$ should be minimized to maximize luminosity. 

\begin{equation}
    \mathcal{L} = \frac{N_1 N_2 f N_b}{4\pi \sigma_x \sigma_y}
    \label{eq:lumi}
\end{equation}

where $N_{1,2}$ is the number of particles contained by each beam, $f$ is the revolution frequency \cite{herr2006concept}.

\begin{figure}[h]
    \includegraphics[width=\columnwidth]{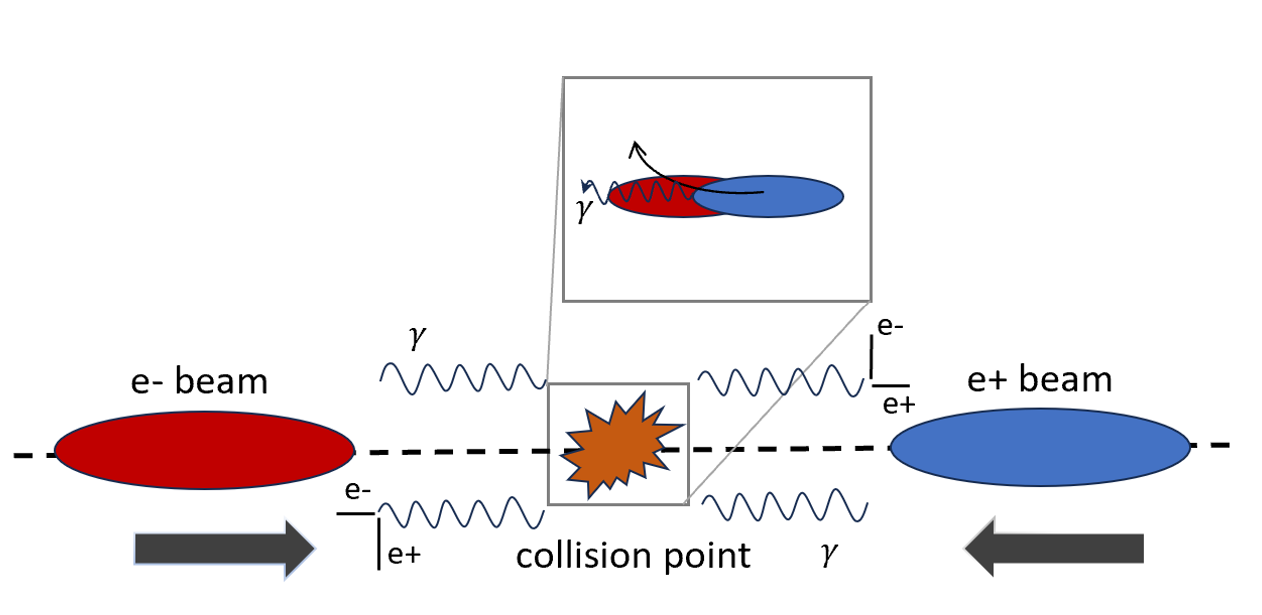}
    \caption{Collision of electron and positron beams, resulting in photon and pair production  at the interaction point.}
    \label{fig:interaction_diagram}
\end{figure}

However, small transverse beam sizes lead to an intense electromagnetic (EM) field \cite{schulte1999beam}, which deflects the charged particles and causes them to emit radiation (Fig. \ref{fig:interaction_diagram}). The intensity of these interactions is characterized by the beamstrahlung parameter $\Upsilon$ in Eq. \eqref{eq:upsilon}. For Gaussian beams, the average and maximum $\Upsilon$ are estimated with Eq. \eqref{eq:upsilon_estimate}.

\begin{equation}
    \Upsilon = \frac{2}{3} \frac{\hbar \omega_c}{E_0}
\label{eq:upsilon}
\end{equation}

\begin{equation}
    \langle \Upsilon \rangle \sim \frac{5}{6}\frac{N r_e^2 \gamma}{\alpha \sigma_z (\sigma_x + \sigma_y) }\sim \frac{5}{12} \Upsilon_\text{max}
    \label{eq:upsilon_estimate}
\end{equation}

where $\omega_c$ is the critical frequency defined at half power spectrum, $E_0$ is the particle energy, $\gamma$ is the relativistic factor, $\alpha$ is the fine structure constant and $r_e$ is the electron's classical radius. 

\begin{figure}[h]
    \includegraphics[width=\columnwidth]{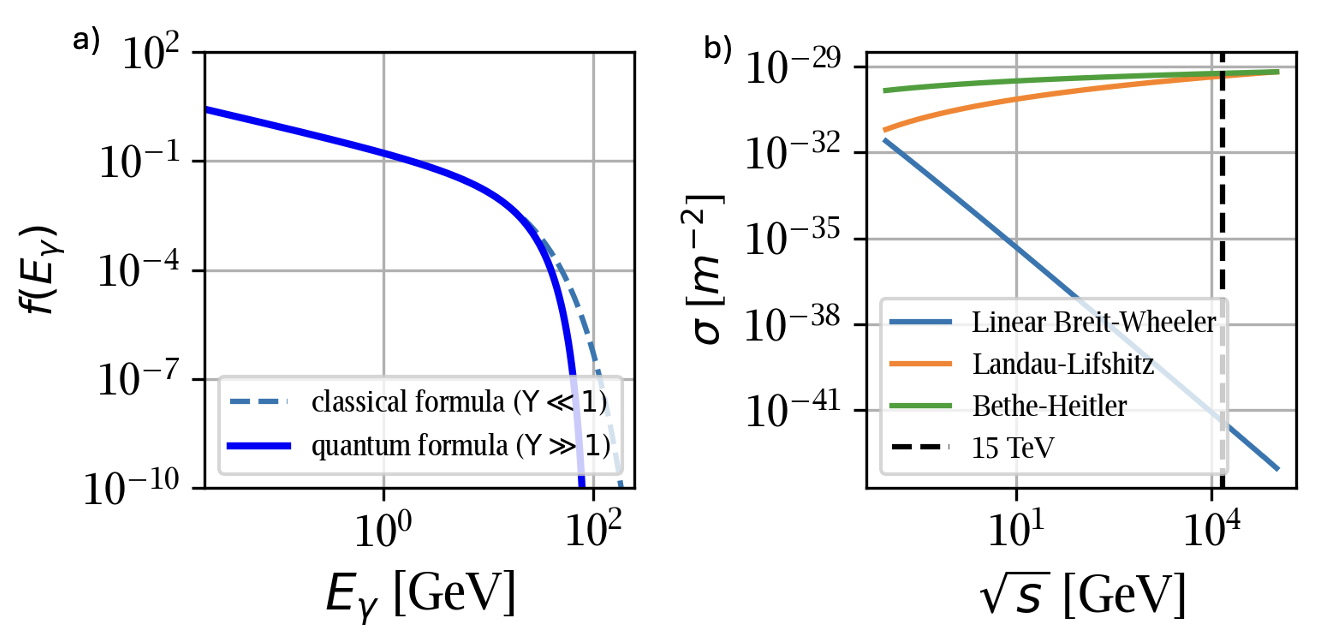}
    \caption{a) Normalized beamstrahlung energy spectrum for ILC-250 GeV assuming no energy spread and constant relativistic magnetic field approximated by the infinite thin wire formula $B = \gamma  \mu_0 I / \left(2\pi \sqrt{\sigma_x^2 + \sigma_y^2}\right)$ $\sim 10^3$ Tesla. b) Cross-section of different incoherent processes as a function of the center of mass energy.}
    \label{fig:quantum_sketch}
\end{figure}

For $\Upsilon \gg 1$, the photon spectrum is given by the Sokolov-Ternov formula, which truncates the photon energy at $E_\gamma = E_0$ as opposed to the classical formula which extends infinitely \cite{sokolov1968synchrotron} (Fig. \ref{fig:quantum_sketch}a). 

\begin{equation}
    \frac{dN_\gamma}{d\bar{x}} = \frac{\alpha}{\sqrt{3}\pi \gamma^2} \left[ \frac{\hbar \omega}{E} \frac{\hbar \omega}{E - \hbar \omega} K_{2/3}(\bar{x}) + \int_{\bar{x}}^{\infty} K_{5/3}(x') dx' \right ]
    \label{eq: quantum spectra}
\end{equation}

where $\bar{x} = \omega/\omega_c \cdot E_0/(E_0- \hbar \omega)  \propto 1/\Upsilon $ is the modified frequency ratio, and $K_{i}$ is the modified Bessel function of order $i$.

The emitted photons may interact with the macroscopic field or collide with other high-energy photons to form electron-positron pairs. The coherent pair production process becomes relevant towards $\Upsilon \sim 10^{3}$, and the incoherent processes are characterized by their cross sections \cite{schulte1999beam} as shown in Fig. \ref{fig:quantum_sketch}b).  

\section{Modelling Approach}

\subsection{Overview of simulation codes}

WarpX represents plasma with macro-particles with EM field simulated on a grid \cite{fedeli2022pushing}. At each timestep $t_n$, the macro-particles are pushed based on the interpolated EM fields, which generate a self-consistent current. Each macro-particle is assigned an optical depth $\tau_n$ updated at each $t_n$ to account for QED effects. When $\tau_n < 0$, quantum events such as beamstrahlung emission are triggered. Here, we used the spectral solver with open boundary conditions, which is more efficient than the typical iterative electrostatic solver. Similarly, GUINEA-PIG uses macro-particles to represent beams, which  are divided into slices since the fields are orientated transversely in the relativistic limit \cite{schulte1999beam}. GUINEA-PIG iterates through all the slices, and the ones at the same $z$ position are allowed to interact. The slices are further divided into cells, where the charges are first distributed on the grid. The field is computed through a Fourier transform method, allowing the particles' positions to be updated accordingly.   

Table \ref{tab:physics-processes} gives an overview of the QED processes, where beamstrahlung and coherent pair production are available in both codes. GUINEA-PIG implements these two processes through rejection sampling of the distribution shown in Eq. \eqref{eq: quantum spectra}.  WarpX interpolates logarithmic look-up tables generated separately from the simulation, where a range of $\Upsilon$ values are specified based on Eq. \eqref{eq:upsilon_estimate}.

\begin{table}[h!b]
  \setlength\tabcolsep{3.5pt} 
  \centering
  \caption{Overview of Physics Processes in Simulation Codes}
  \label{tab:physics-processes}
  \begin{tabular}{lcc}
    \toprule
    \textbf{Physics Process}       & \textbf{GUINEA-PIG} & \textbf{WarpX} \\
    \midrule
    Beamstrahlung                & $\checkmark$        & $\checkmark$   \\
    Bethe-Heitler                 & $\checkmark$        & Soon  \\
    Linear Breit-Wheeler          & $\checkmark$        & Soon       \\
    Landau-Lifshitz               & $\checkmark$        & Soon      \\
    Coherent Pair Production      & $\checkmark$        & $\checkmark$       \\
    Trident Cascade               & $\checkmark$        & $\times$       \\
    Hadronic Production + Minijets& $\times$            & $\times$       \\
    Electron-Laser Interaction    & $\times$            & $\checkmark$   \\
    \bottomrule
  \end{tabular}
\end{table}

\subsection{Test cases}

\begin{figure}[h]
    \includegraphics[width=\columnwidth]{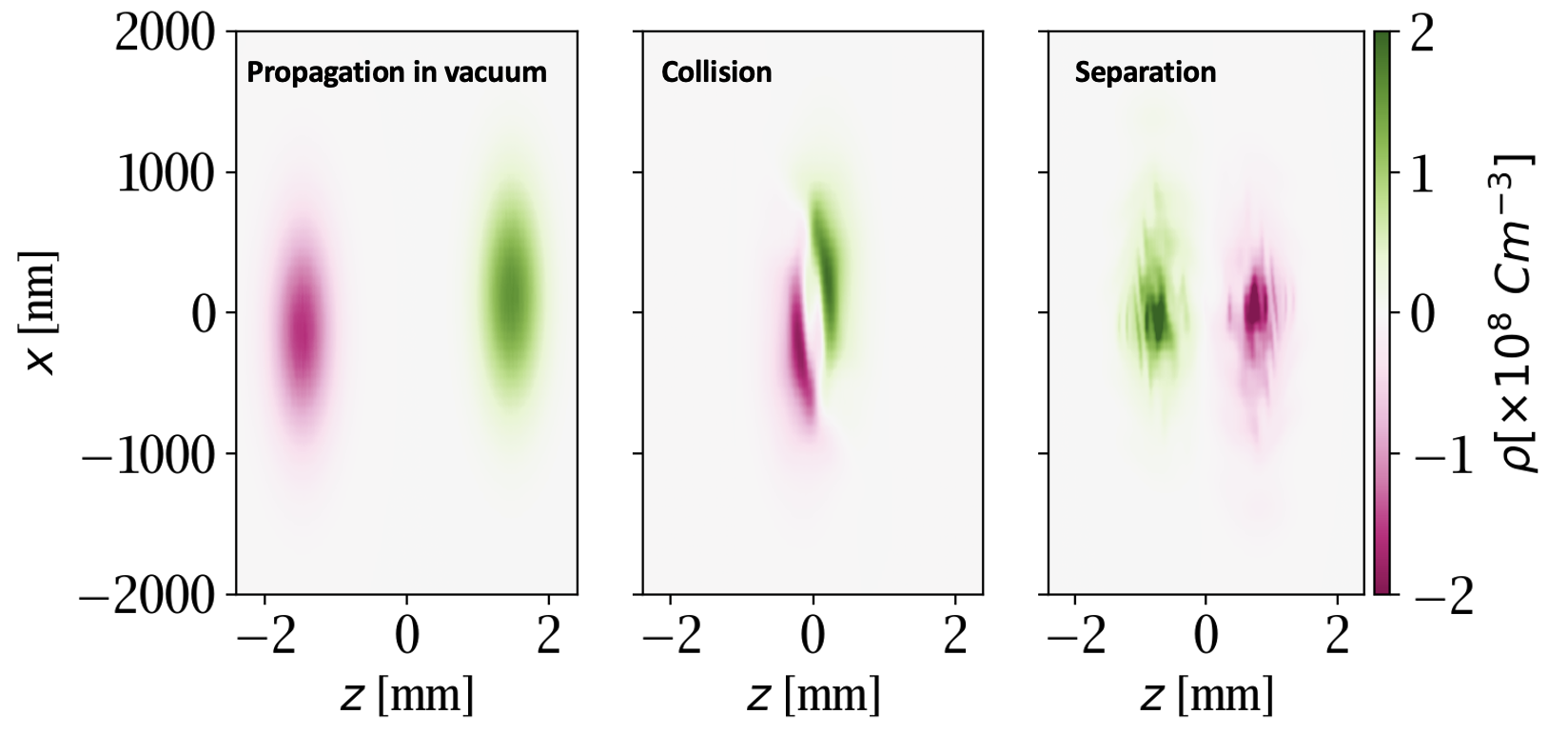}
    \caption{Visualization of nominal ILC-250 GeV  beam-beam collision in WarpX. Note that the aspect ratio is not correct.}
    \label{fig:setup}
\end{figure}

The test cases conducted include the designs of the ILC \cite{bambade2019international} and the ultratight collider \cite{yakimenko2019prospect}. In both codes, mono-energetic Gaussian beams are initialized with equal energies. The beams are initially separated in the z-direction then allowed to propagate and interact as depicted in Fig. \ref{fig:setup}.

For the test cases of beamstrahlung radiation and luminosity, we consider ILC scenarios with $\sqrt{s} =$ 250 GeV, 500 GeV, 1 TeV. To benchmark the coherent pair production, we  consider an ultra-tight spherical beam collision configuration with $\sigma_x = \sigma_y = \sigma_z = 10$ nm \cite{yakimenko2019prospect}, where $\Upsilon$ reaches approximately $10^3$. Convergence tests were performed to ensure validity of the results. 

\section{Results}

\subsection{Physics comparison}

 The luminosity results for nominal and disrupted ILC scenarios shows an agreement within 5\% (Table \ref{tab:lumi}) . 
 With an offset $\Delta x = \sigma_x/2$ in the $x$ direction, the luminosity decreases by a factor of  $\Delta \mathcal{L} / \mathcal{L} = \exp{(-\Delta x ^2 / \sigma_x^2)} \sim 0.9$, which conforms with geometric predictions \cite{herr2006concept}.

\begin{table}[h]
  \centering
  \caption{ILC-based luminosity with and without an offset of $\Delta x = \sigma_x/2$. GP stands for GUINEA-PIG.}
  \label{tab:lumi}
  \begin{tabular}{cccccc}
    \toprule
    & & \multicolumn{2}{c}{$\mathcal{L}_{\Delta x = 0}$ [$\mathrm{m^{-2}}$]} & \multicolumn{2}{c}{$\mathcal{L}_{\Delta x = \sigma_x /2}$ [$\mathrm{m^{-2}}$]} \\
    \cmidrule(lr){3-4} \cmidrule(lr){5-6}
    $\sqrt{s}$ [GeV] & & WarpX & GP & WarpX & GP \\
    \midrule
    250 & & 1.81 & 1.89 & 1.70 & 1.71 \\
    500 & & 2.45 & 2.57 & 2.29 & 2.38 \\
    1000  & & 5.83 & 5.93  & 5.46 & 5.46 \\
    \bottomrule
  \end{tabular}
\end{table}

\begin{figure}[h]
\includegraphics[width=\columnwidth]{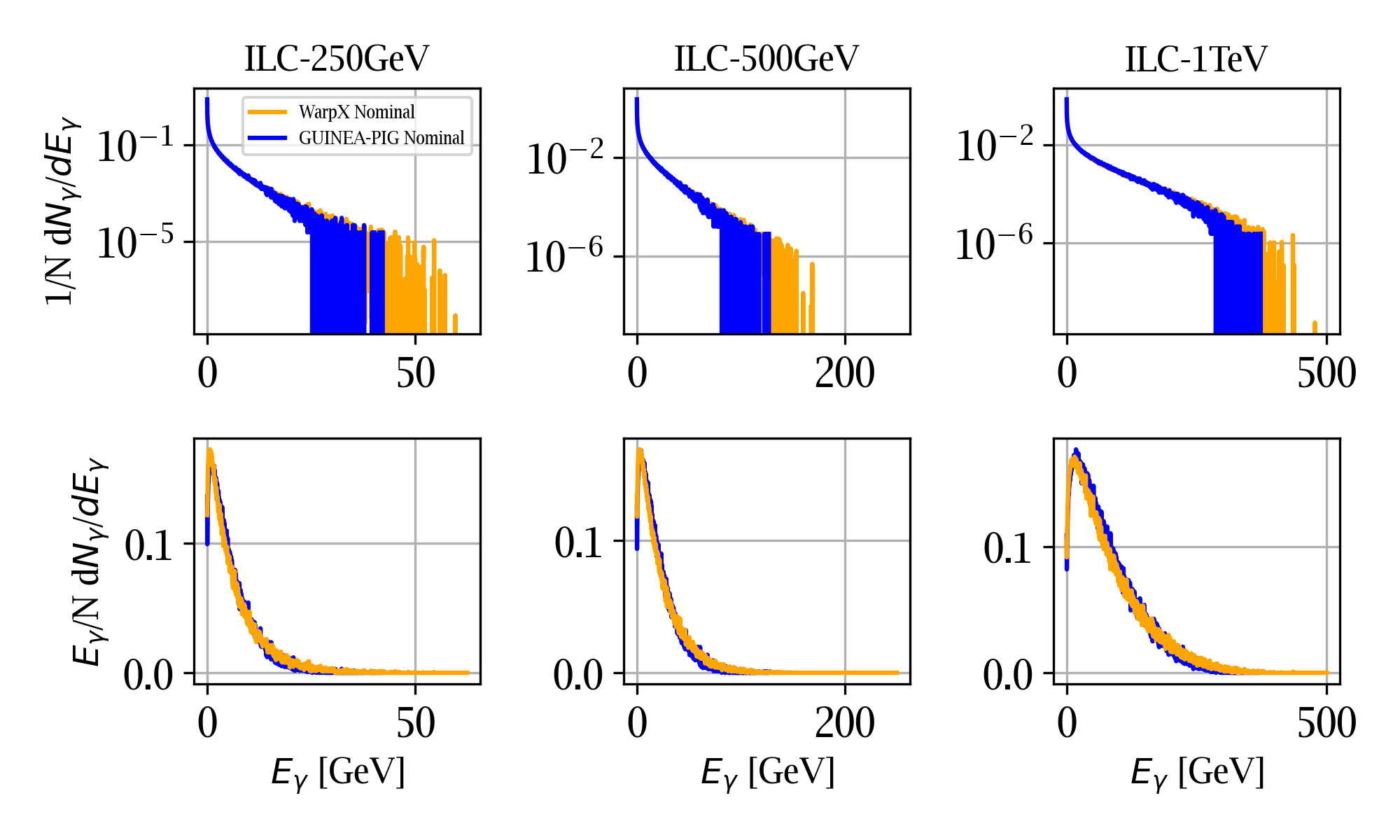}
\caption{Photon spectra for nominal ILC scenarios. The top row shows normalized distributions, and the bottom row shows distributions normalized by $E_\gamma$. Spectra with the offset are very similar and are omitted to reduce clutter.}
\label{fig:photon_spectrum}
\end{figure}

\begin{figure}[h]
\includegraphics[width=\columnwidth]{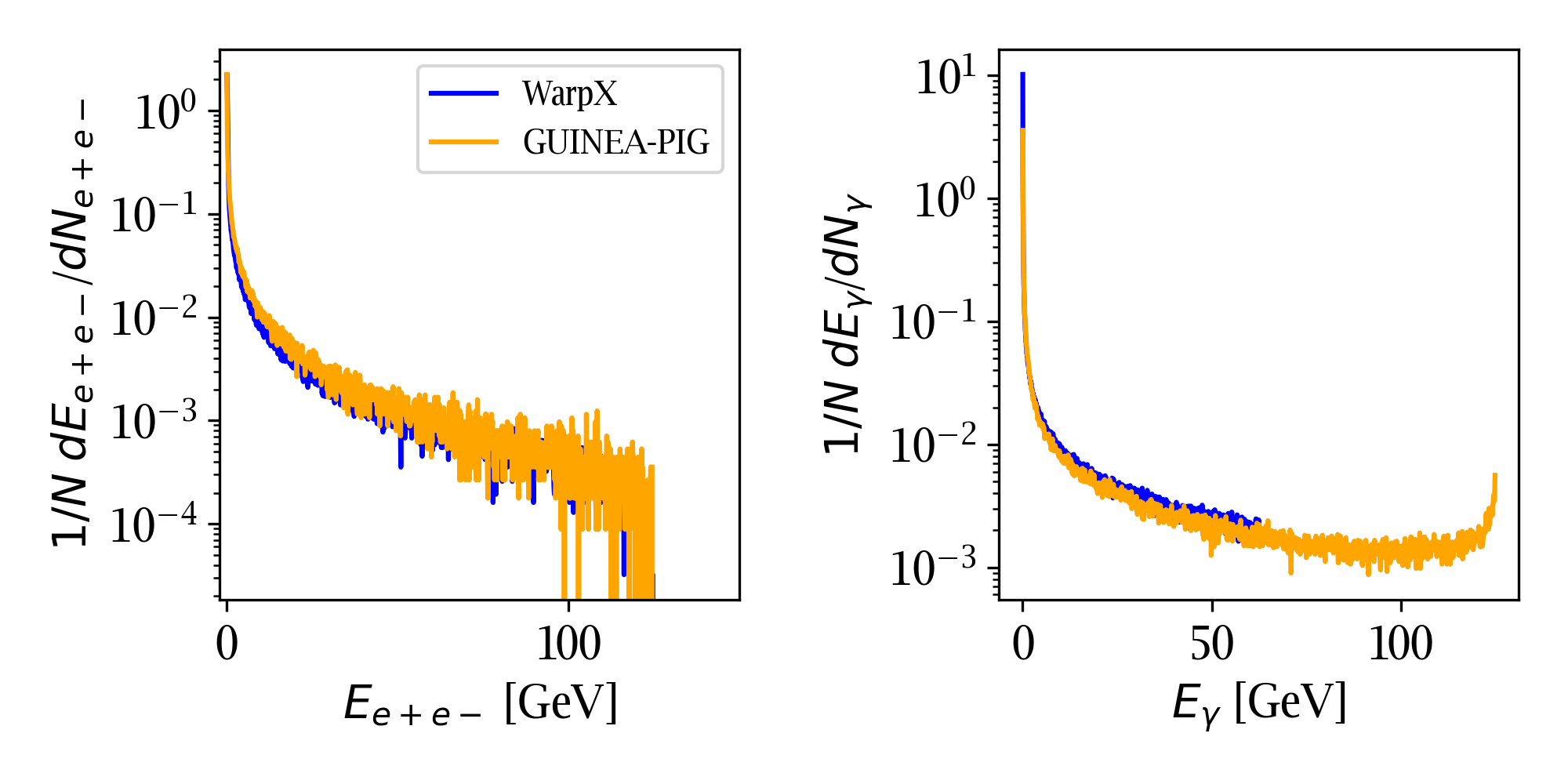}
\caption{Normalized spectrum of coherent electron-positron pairs and photon produced in the ultratight scenario.}
\label{fig:ultratight}
\end{figure}

Figure \ref{fig:photon_spectrum} shows the benchmarked photon spectrum for nominal ILC scenarios at $\sqrt{s} = $ 250 GeV, 500 GeV and 1 TeV. Here, $10^{5}$ macro-particles are used with a grid size $(n_x, n_y, n_z)$ of $128 \times 128 \times 32$.  Diagnostic-wise, it is important to have a sufficient number since the photon distribution is skewed towards the lower energy. For both codes, the maximum beamstrahlung parameter $\Upsilon \lesssim 1$ leads to no coherent pair produced as expected. In addition, the benchmarked spectrum for coherent pairs and beamstrahlung of the ultratight collider scenario indicates solid agreement (Fig. \ref{fig:ultratight}). 
  
\newpage
\subsection{Performance comparison}

\begin{figure}[h]
    \includegraphics[width=\columnwidth]{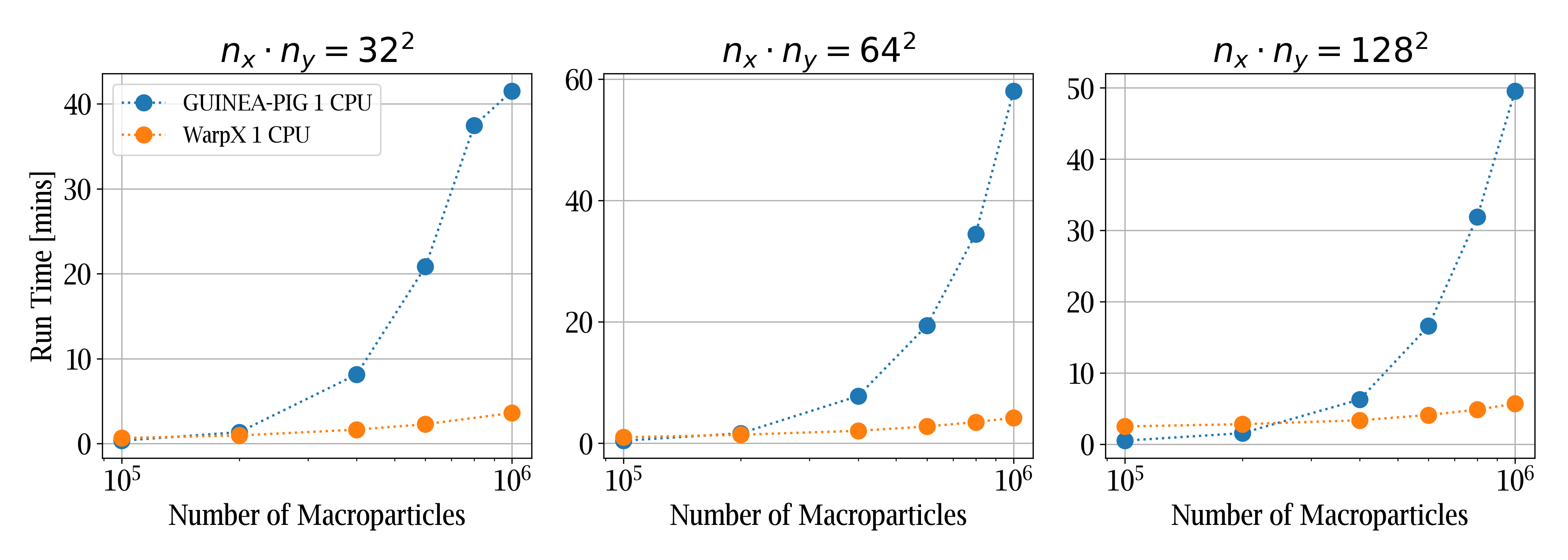}
    \caption{Performance scaling using 1 CPU for increasing number of macro-particles with different fixed grid sizes. }
    \label{fig:1cpu}
\end{figure}

First, we performed a fair comparison with 1 CPU between WarpX and GUINEA-PIG. The performance scaling for both codes as a function of  macro-particles number $n_p$ is shown in Fig. \ref{fig:1cpu}. GUINEA-PIG run time increases exponentially with increasing $n_p$ while WarpX run time increases approximately linearly.  

\begin{figure}[h]
    \includegraphics[width=\columnwidth]{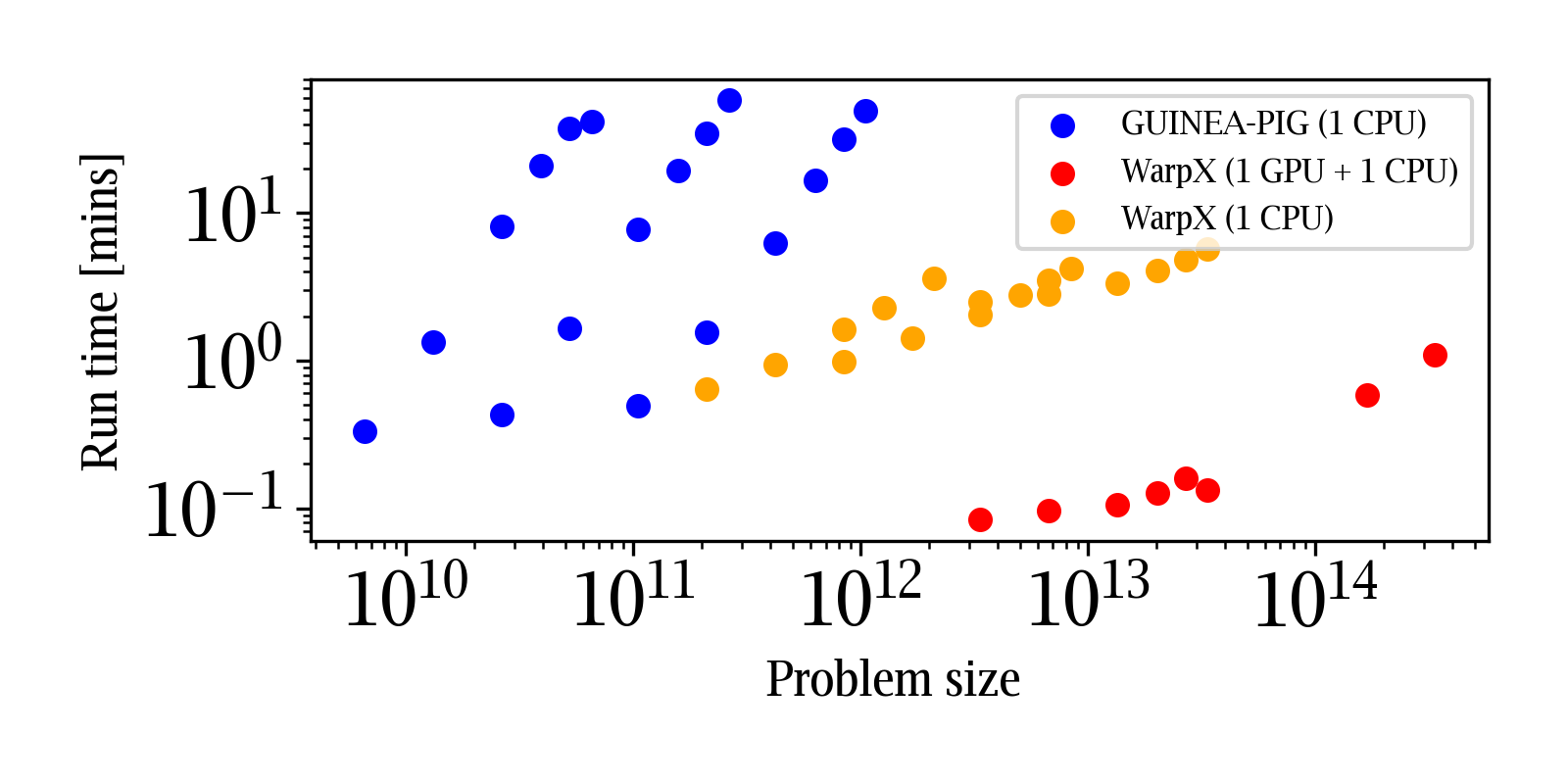}
    \caption{Performance scaling of WarpX and GUINEA-PIG as a function of problem size defined as $n_x n_y n_z n_t n_m$. }
    \label{fig:figure_of_merit}
\end{figure}

Next, we considered GPU acceleration with WarpX, which leads to multiple order of magnitudes speed-up (Fig. \ref{fig:figure_of_merit}). With modern HPCs, WarpX is often run with multiple CPUs and GPUs, which can accelerate the simulation, and enable high resolution beam-beam studies. 

\section{CONCLUSION}

Using the ILC and ultratight collision scenarios, we established excellent agreement the mutual physical processes of WarpX and GUINEA-PIG. The performance comparison indicates a significant speed advantage that WarpX offers in both serial and parallelized runs. This is crucial in advancing the capabilities of simulation tools to the high-energy and particle physics community. Future work could involve benchmarking the soon-to-be-implemented incoherent pair production processes and comparing further physics outputs such as luminosity-energy spectra, spatial field, and charge density.

\newpage

%
%

\ifboolexpr{bool{jacowbiblatex}}%
	{\printbibliography}%
	{%
	
	
} 
%
%


\end{document}